\documentclass[pra,aps,twocolumn,showpacs,superscriptaddress]{revtex4}
\usepackage{graphicx,latexsym}
\usepackage{amsmath}
\newcommand{\sect}[2]{{\par\it #1.---}{#2}}
\newcommand{\beq}{\begin{equation}}
\newcommand{\eeq}{\end{equation}}
\newcommand{\bqn}{\begin{eqnarray}}
\newcommand{\eqn}{\end{eqnarray}}
\newcommand{\bqns}{\begin{eqnarray*}}
\newcommand{\eqns}{\end{eqnarray*}}
\newcommand{\bary}{\begin{array}}
\newcommand{\eary}{\end{array}}
\newcommand{\non}{\nonumber}

\begin{document}
\title{Exact approaches to charged particle motion in a
time-dependent flux-driven ring}

\author{Pi-Gang Luan}
\affiliation{Institute of Optical Sciences, National Central
University, Chung-Li 32054, Taiwan}
\author{Chi-Shung Tang}
\affiliation{Physics Division, National Center for Theoretical
Sciences, P.O. Box 2-131, Hsinchu 30013, Taiwan}
\date{\today}

\begin{abstract}
We consider a charged particle which is driven by a time-dependent
flux threading a circular ring system.  Various approaches including
classical treatment, Fourier expansion method, time-evolution
method, and Lewis-Riesenfeld method are used and compared to solve
the time-dependent problem.  By properly managing the boundary
condition of the system, a time-dependent wave function of the
charged particle can be obtained by using a non-Hermitian
time-dependent invariant, which is a specific linear combination of
initial angular-momentum and azimuthal-angle operators. The
eigenfunction of the linear invariant can be realized as a
Gaussian-type wave packet with a peak moving along the
classical angular trajectory, while the distribution of the wave
packet is determined by the ratio of the coefficient of the initial
angle to that of the initial canonical angular momentum. In this
topologically nontrivial system, we find that although the classical
trajectory and angular momentum can determine the motion of the wave packet;
however, the peak position is no longer
an expectation value of the angle operator. Therefore, in such a
system, the Ehrenfest theorem is not directly applicable.
\end{abstract}

 \maketitle

\section{Introduction}

Charged particle driven by a time-dependent perturbation in a
quantum system is a nontrivial fundamental
issue~\cite{Wees91,FF96,TC99,LC00,GBG,MG04,LT05}. One can access the
charged particle wave function by placing it in a quantum ring
threaded by a time-dependent magnetic flux.  The vector potential
${\bf A}(t)$ associated with the time-dependent flux $\Phi(t)$ times
the charge $q$ leading to a phase shift proportional to the number
of flux quanta penetrating the ring, this is known as Aharonov-Bohm
(AB) effect~\cite{AB59,AS87,CP94}.  In adiabatic cyclic evolution,
Berry~\cite{Berry84} first discovered that there exists a geometric
phase.  Later on, Aharonov and Anandan (AA) removed the adiabatic
restriction to explore the geometric phase for any cyclic
evolution~\cite{AA87}.  Time-dependent fields are also used to deal
with field driven Zener tunneling, in which nonadiabaticity plays a
crucial role~\cite{LB85,GT87,MB89}.

In mesoscopic systems, a number of manifestations of the AB effect
have been predicted and
verified~\cite{B83L90,AB91-98,AB-Nature,BK03}.  On the other hand,
Stern demonstrated that the Berry phase affects the particle motion
in the ring similar to the AB effect, and a time-dependent Berry
phase induces a motive force~\cite{Stern92}.  Recently, AA phases
have been found to play the role of classical canonical actions and
are conserved in the adiabatic evolution of
noneigenstates~\cite{Niu03}.

In the present work, we consider a noninteracting spinless charged
particle moving cyclically in a quantum ring in the presence of a
time-dependent vector potential. Such a particle motion can be
described by the time-dependent Schr{\"o}dinger equation
\begin{equation}
i\hbar\frac{\partial\psi}{\partial
t}=\hat{H}(t)\psi\label{firsteq}\, ,
\end{equation}
where the Hamiltonian $\hat{H}(t)$ is induced by an external
time-dependent vector potential ${\bf A}(t)$, given by
\bqn
\hat{H}(t)&=&\frac{1}{2m}\left[\hat{{\bf P}}-q{\bf A}(t)\right]^2\non\\
&=&\frac{1}{2I}\left[\hat{L}-qRA(t)\right]^2. \label{HH}\eqn
Here $\hat{{\bf P}}= {\bf e}_\theta\hat{P}_\theta$ is the canonical
momentum operator with ${\bf e}_\theta$ being the unit vector along
the azimuthal angle $\theta$; $\hat{L}=\hat{L}_z=(\hat{\bf r}\times
\hat{\bf P})_z$ is the canonical angular momentum operator in the
$z$ direction; $I=mR^2$ is the moment of inertia of the particle;
${\bf A}(t)=A(t){\bf e}_\theta$ is the vector potential; and $R$ is
the radius of the circular ring.  This time-dependent dynamical
problem can be solved by taking into account the Fourier expansion,
time evolution operator, and Lewis-Riesenfeld (LR)
method~\cite{Lewis,LR}.

\section{A Classical treatment}\label{classical}

We first analyze the time-dependent problem in a classical manner.
The time-varying magnetic flux induces an electric field ${\bf
E}=E{\bf e}_\theta$ such that $E=-\partial A/\partial t$. The
charged particle thus obtain a kinetic momentum increment during the
time interval from $0$ to $t$, namely
\begin{equation}
\Delta p_c=\Delta
(mv)=m[v(t)-v(0)]=-q\left[A(t)-A(0)\right],\label{eq1}
\end{equation}
where $p_c=mv$ is the kinetic momentum. It should be noted that both
$p_c$ and $qA$ are not conservative quantities, while from
Eq.~(\ref{eq1}) we see that the canonical momentum $P_c$ is a
constant of motion:
\begin{equation}
P_c(t)=mv(t)+qA(t)=mv(0)+qA(0)=P_c(0). \label{eq2}
\end{equation}

Comparing the two identities in Eq.~(\ref{HH}), we see that the
result of Eq.~(\ref{eq1}) is equivalent to
\begin{equation}
\Delta
l_c=I\left[\omega(t)-\omega(0)\right]=-\frac{q}{2\pi}\left[\Phi(t)-\Phi(0)\right],
\end{equation}
where $l_c=({\bf r}\times {\bf p}_c)_z=I\omega$ indicates the
kinetic angular momentum, $\omega$ is the angular velocity,  $\Phi$
is the magnetic flux threading the ring, and the fact
\begin{equation}
\Phi(t)=2\pi\,RA(t)
\end{equation}
has been used.
Also, from Eq.(\ref{eq2}) we have
\begin{equation}
L_c(t)=l_c(t)+\frac{q}{2\pi}\Phi(t)=l_c(0)+\frac{q}{2\pi}\Phi(0)=L_c(0).
\end{equation}
Thus the {\it canonical angular momentum} $L_c=({\bf r}\times {\bf P}_c)_z$ is
also a constant of motion.

Now we define the {\it writhing number} as
\begin{equation}
n_\Phi(t)\equiv\frac{\Phi(t)}{\Phi_0},
\end{equation}
where $\Phi_0= h/q$ is a flux quantum. We also define
\begin{equation}
L_c\equiv n_0\hbar,\;\;\;\;l_c(t)\equiv n_c(t)\hbar,
\end{equation}
then we have
\begin{equation}
n_c(t)=n_0-n_\Phi(t).
\end{equation}
All of these $n$'s are {\it real} numbers.

Now the angular position of the driven particle is given by \bqn
\theta_c(t)&=&\theta_0+\int^t_0\frac{n_c(\tau)\hbar}{I}d\tau\non\\
&=&\theta_0+\omega_0t-\int^t_0\frac{n_\Phi(\tau)\hbar}{I}d\tau. \eqn
Here $\theta_0$ indicates the initial azimuthal angle; and
$\omega_0=\omega(0)=n_0\hbar/I$ stands for the initial angular
velocity. Below we denote the initial kinetic angular momentum $l_0
\equiv l_c(0)$ for simplicity.

\section{A Fourier expansion method}

The simplest method for solving the time-dependent flux-driven
problem is the Fourier expansion method. The first thing about the
system we discuss is that the wave function satisfies the periodic
boundary condition:
\begin{equation}
\psi(\theta,t)=\psi(\theta+2\pi,t).\label{prdbc}
\end{equation}
The most general form of $\psi$ for the present problem is thus written as
\begin{equation}
\psi(\theta,t)=\sum^{\infty}_{n=-\infty}c_nf_n(t)e^{in\theta},\label{expansion}
\end{equation}
where the $c_n$'s are appropriate coefficients to be determined by
the initial and the boundary conditions.

Substituting Eq.~(\ref{expansion}) into Eq.~(\ref{firsteq}), we can
find \bqn
&&\sum^{\infty}_{n=-\infty}i\hbar\,c_n \dot{f}_n(t) e^{in\theta}\non\\
&&=\sum^{\infty}_{n=-\infty}c_nf_n(t)\frac{\left(n\hbar-qRA(t)\right)^2}{2I}e^{in\theta}.
\label{feq}
\eqn

Solving Eq.(\ref{feq}), after some procedures we obtain
\begin{equation}
f_n(t)=\exp\left\{-\frac{i}{2I\hbar}\int^{t}_{0}\left[n\hbar-qRA(t)\right]^2dt\right\},
\end{equation}
and thus
\begin{equation}
\psi(\theta,t)=\sum^{\infty}_{n=-\infty}c_n\,
\exp\left\{-\frac{i\hbar}{2I}\int^{t}_{0}\left[n-n_\Phi(\tau)\right]^2d\tau+in\theta\right\}.\label{longtime}
\end{equation}

As an simple example, let us choose
\begin{equation}
c_n=N \exp\left[-\sigma^2(n-n_0)^2-i \theta_0(n-n_0)\right],\label{cn}
\end{equation}
where $N$ indicates an appropriate normalization constant; $\sigma$, $\theta_0$ and $n_0$
are real numbers, and $n$ is an integer.

Substituting Eq.(\ref{cn}) into Eq.(\ref{longtime}),
we get
\bqn
&&\psi(\theta,t)=N
\exp\left[-\frac{i}{\hbar}\int^t_0\frac{l_c^2(\tau)}{2I}d\tau+in_0\theta_c(t)\right]
\label{pw}\\
&&\times\sum^{\infty}_{n=-\infty}
\exp\left\{-\sigma^2\left(1+\frac{it}{T}\right)(n-n_0)^2+in\left[\theta-\theta_c(t)\right]
\right\},\non \eqn where $T=2I\sigma^2/\hbar$.  Applying the Poisson
summation formula
\begin{equation}
\sum^{\infty}_{n=-\infty}f(n)=
\sum^{\infty}_{n=-\infty}\left(\int^{\infty}_{-\infty}f(x)e^{i2\pi n
x}dx\right)
\end{equation}
on the function \bqn
f(x)=\exp{\left[-\sigma^2\left(1+\frac{it}{T}\right)(x-n_0)^2
+i\left(\theta-\theta_c(t)\right)x\right]},\non \\ \eqn we can obtain
an alternative expression \bqn
&&\psi(\theta,t) =N\sqrt{\frac{\pi}{\sigma^2(1+\frac{it}{T})}}
\exp\left[-\frac{i}{\hbar}\int^t_0\frac{l_c^2(\tau)}{2I}d\tau\right]\label{pkt}\\
&&\times\sum^{\infty}_{n=-\infty}
\exp\left[-\frac{\left(\theta-\theta_c(t)+2n\pi\right)^2}
{4\sigma^2\left(1+\frac{it}{T}\right)}+in_0(\theta+2n\pi)\right].\non
\eqn

\section{A Time evolution method}

In this section, we shall present how to get the general solution
shown in the previous section in terms of the time evolution
operator $\hat{U}(t)$. The state $|\psi(t)\rangle$ is connected with
the initial state $|\psi(0)\rangle$ through
\begin{equation}
|\psi(t)\rangle=\hat{U}(t)|\psi(0)\rangle.
\end{equation}
and the wave function $\psi(\theta,t)$ is given by
\begin{equation}
\psi(\theta,t)=\langle\theta|\hat{U}(t)|\psi(0)\rangle,
\end{equation}
where $|\theta\rangle$ is the $\theta$-eigenket in the
Schr\"{o}dinger picture, which will be explained later.

To begin with, we introduce the canonical commutator
\begin{equation} [\hat{\theta}(0),\hat{L}(0)]=i\hbar. \end{equation}
From this identity, we have
\begin{equation}
[\hat{\theta}(t),\hat{L}(t)]
=\hat{U}^\dagger(t)[\hat{\theta}(0),\hat{L}(0)]\hat{U}(t)=i\hbar.\label{heis}
\end{equation}

Utilizing Eq.(\ref{heis}) we can derive
\begin{equation}
\frac{d\hat{L}(t)}{dt}=\frac{[\hat{L}(t),\hat{H}(t)]}{i\hbar}=0,
\end{equation}
 we thus obtain the identity $\hat{L}(t)=\hat{L}(0).$ Following
 similar procedure it is easy to obtain
\begin{equation}
\frac{d\hat{\theta}(t)}{dt}=\frac{[\hat{\theta}(t),\hat{H}(t)]}{i\hbar}
=\frac{\hat{L}(0)-n_\Phi(t)\hbar}{I},
\end{equation}
which gives us
\begin{equation}
\hat{\theta}(t)=\hat{\theta}(0)+\frac{\hat{L}(0)t}{I}-\int^t_0\frac{n_\Phi(\tau)\hbar
}{I}d\tau.\label{theta}
\end{equation}
We can see that the canonical angular momentum
is a constant of motion. This is consistent with the classical
results discussed in Sec.~\ref{classical}.

From the above results we have
\begin{equation}
[\hat{H}(t),\hat{H}(t')]=0
\end{equation}
for any two times $t$ and $t'$. Hence the time evolution operator is
simply given by \bqn
\hat{U}(t)&=&\exp\left[-\frac{i}{\hbar}\int^t_0\hat{H}(\tau)d\tau\right]\non\\
&=&\exp\left[-\frac{i\hbar}{2I}\int^t_0
\left(\frac{\hat{L}(0)}{\hbar}-n_\Phi(\tau)\right)^2 d\tau\right].
\eqn

To proceed further, we define $|n\rangle$ as the eigenket of
$\hat{L}(0)$ obeying
\begin{equation}
\hat{L}(0)|n\rangle=n\hbar |n\rangle.
\end{equation}
In addition,  we also assume that $|\theta\rangle$ is an eigenket of
$e^{i\hat{\theta}(0)}$ obeying
\begin{equation}
e^{i\hat{\theta}(0)}|\theta\rangle=e^{i\theta}|\theta\rangle.
\end{equation}
The orthogonal conditions of the two eigenkets are given by
\begin{equation}
\langle m|n \rangle =
\frac{1}{2\pi}\int^{2\pi}_0e^{i(m-n)\theta}d\theta=\delta_{mn}
\end{equation}
and
\begin{equation}
\langle\theta|\theta' \rangle =
\frac{1}{2\pi}\sum^{\infty}_{n=-\infty}e^{in(\theta-\theta')}=\delta(\theta-\theta'),
\end{equation}
which can be easily derived from the closure relations
\begin{equation}
\sum^{\infty}_{n=-\infty}|n\rangle\langle n|=1,\;\;\;\;\;
\int^{2\pi}_0d\theta\, |\theta\rangle\langle\theta|=1
\end{equation}
and taking into account the definition
\begin{equation}
\langle\theta|n\rangle=\frac{1}{\sqrt{2\pi}}e^{in\theta}=\langle n|\theta\rangle^*.
\end{equation}

We note in passing that both $\theta$ and $\theta'$ are defined in
the interval $[0,2\pi)$. In the coordinate representation, the
$e^{in \theta}$ is an eigenfunction of $\hat{L}_{\rm
rep}=-i\hbar\partial/\partial\theta$ with corresponding eigenvalue
$n\hbar$.  This result can be expressed as
\begin{equation}
\langle\theta|\hat{L}(0)|n\rangle=\hat{L}_{\rm
rep}\langle\theta|n\rangle = n\hbar \langle\theta|n\rangle.
\label{leigen1}
\end{equation}
The wave function $\psi$ now can be calculated: \bqn
&&\psi(\theta,t)=\sum^{\infty}_{n=-\infty}\langle \theta
|\hat{U}(t)|n\rangle\langle n
|\psi(0)\rangle\non\\
&&=\sum^{\infty}_{n=-\infty}\frac{\langle n
|\psi(0)\rangle}{\sqrt{2\pi}} e^{-\frac{i\hbar}{2I}\int^t_0
\left[n-n_\Phi(\tau)\right]^2 d\tau+in\theta}.\label{time}
\eqn
If we define
\begin{equation}
c_n=\frac{\langle n
|\psi(0)\rangle}{\sqrt{2\pi}},
\end{equation}
then the result of Eq.(\ref{time}) becomes that of Eq.(\ref{longtime}).

\section{A Lewis-Riesenfeld method}

In this section, we briefly review the LR method and then apply it
to solve the present problem. We shall show that LR method is not
directly applicable, however, a simple modification concerning about
the boundary condition makes it applicable to solving the problems
with periodic boundary condition.

Traditionally, to utilize the LR method~\cite{LR} solving a
time-dependent system, we have to find an operator $\hat{Q}(t)$ such
that
\begin{equation}
i\hbar\frac{d \hat{Q}}{dt}=i\hbar\frac{\partial \hat{Q}}{\partial
t}+[\hat{Q},\hat{H}]=0, \label{invariant},
\end{equation}
and then find its eigenfunction $\varphi_\lambda(\theta,t)$ satisfying
\begin{equation}
\hat{Q}(t)\,\varphi_{\lambda}(\theta,t)=\lambda\,\varphi_{\lambda}(\theta,t),\label{eigenphi}
\end{equation}
with $\lambda$ being the corresponding eigenvalue.
A wave function $\psi_\lambda (\theta,t)$ satisfying Eq.(\ref{firsteq}) is
then obtained via the relation
\begin{equation} \label{phaseeq}
\psi_\lambda(\theta,t)=e^{i\alpha_{\lambda}(t)}\,\varphi_{\lambda}(\theta,t),
\end{equation}
where $\alpha(t)$ is a function of time only, satisfying
\begin{equation}
\dot{\alpha}_{\lambda}=\varphi^{-1}_{\lambda}
(i{\partial}/{\partial
t}-\hat{H}/\hbar)\varphi_{\lambda}.\label{alpha1}
\end{equation}
A general solution $\psi$ of Eq.(\ref{firsteq}) is then given by
\begin{equation}
\psi(\theta,t)={\sum_\lambda} g(\lambda)\psi_\lambda(\theta,t),
\end{equation}
where $g(\lambda)$ is a weight function for $\lambda$.

To proceed let us assume the time-dependent invariant operator
$\hat{Q}(t)$ takes the linear form~\cite{GBG}
\begin{equation}
\hat{Q}(t)=a(t)\hat{L}+b(t)\hat{\theta}+c(t)\, ,
\label{iabc}
\end{equation}
in which $a(t)$, $b(t)$, and $c(t)$ are time-dependent $c$-number
functions to be determined.

Substituting Eq.(\ref{iabc}) into
Eq.(\ref{invariant}) and solving these operator equations, we get
\begin{equation}
a(t)=a_0-\frac{b_0t}{I},\;\;\;\;\;
b(t)=b_0,\label{a0b0}
\end{equation}
\bqn
c(t)&=&c_0+b_0\int^t_0\frac{n_\Phi(\tau)\hbar}{I}d\tau,
\label{c0}
\eqn
where $a_0$, $b_0$, and $c_0$ are arbitrary complex constants.
Furthermore, substituting Eqs.~(\ref{a0b0}) and (\ref{c0}) into Eq.(\ref{iabc}), we find
\begin{equation} \hat{Q}(t)=a_0\,\hat{L}(0)+b_0\,\hat{\theta}(0)+c_0
=\hat{Q}(0).\end{equation} In other words, the invariant $\hat{Q}$
in the Heisenberg picture is precisely the linear combination of the
initial canonical angular momentum $\hat{L}(0)$ and the initial
azimuthal angle $\hat{\theta}(0)$ with an arbitrary constant $c_0$.
Note that in our system the $\hat{L}$ operator is also an invariant.

How does the eigenvalue $\lambda$ evolve in time? Multiplying the factor
$e^{i\alpha(t)}$ on both sides of Eq.(\ref{eigenphi}), we get
\begin{equation}
\hat{Q}(t)\,\psi_{\lambda}(\theta,t)=\lambda\,\psi_{\lambda}(\theta,t).\label{eigenpsi}
\end{equation}
Partially differentiating the both sides of Eq.(\ref{eigenpsi})
with respect to time and using Eq. (\ref{invariant}), we
find
\begin{equation}
\lambda(t)=\lambda(0),\label{lmbd}
\end{equation}
thus $\lambda$ is a constant.

To find a solution of Eq.(\ref{firsteq}), we have to solve Eq.(\ref{eigenphi}) first.
By solving Eq.(\ref{eigenphi}), we get
\begin{equation}
\varphi_{\lambda}(\theta,t)=\exp\left[\frac{i}{\hbar} \left(\mu(t)\theta
-\frac{1}{2}\nu(t)\theta^2 \right)  \right],\label{phisol}
\end{equation}
where
\begin{equation}
\mu(t)=\frac{\lambda-c(t)}{a(t)},\;\;\;\;\;\nu(t)=\frac{b_0}{a(t)}.
\end{equation}
Substituting Eq.~(\ref{phisol}) into Eq.~(\ref{alpha1}), we obtain
\begin{equation}
\alpha_{\lambda}(t)=\alpha_\lambda(0)-\int^t_0
\frac{\left[\eta^2(\tau)+i\hbar\nu(\tau)\right]}{2I\hbar} d\tau,\label{mueta}
\end{equation}
where
\begin{equation}
\eta(\tau)\equiv\mu(\tau)-n_\Phi(\tau)\hbar.
\end{equation}
In deriving Eq.(\ref{mueta}) we have used the two identities:
\begin{equation}
\dot{\mu}=\frac{\nu\left(\mu-n_\Phi\hbar\right)}{I},\;\;\;
\dot{\nu}=\frac{\nu^2}{I}.\label{munu}
\end{equation}
Here we see that in general $\alpha_\lambda(t)$ is a complex function.

Although the form of
$\psi_\lambda(\theta,t)=e^{i\alpha_\lambda(t)}\phi_{\lambda}(\theta,t)$
is indeed a solution of Eq.(\ref{firsteq}), however, it does not
satisfy the periodic boundary condition [see Eq.(\ref{prdbc})]. This
problem can be resolved by defining the total wave function
$\psi(\theta,t)$ as the summation of all
$\psi_\lambda(\theta+2n\pi,t)$ terms: \bqn
&&\psi(\theta,t)=\sum^{\infty}_{n=-\infty}\psi_\lambda(\theta+2n\pi,t)\non\\
&&=\sum^{\infty}_{n=-\infty}
\exp\left[i\alpha_\lambda(t)+\frac{i}{\hbar} \mu(t)(\theta+2n\pi)\right.\non\\
&&\hspace{3cm}\left.-\frac{i}{2\hbar}\nu(t)(\theta+2n\pi)^2 \right].\label{result1}
\eqn

It can also be transformed to the equivalent form below using the
Poisson summation formula: \bqn &&\psi(\theta,t)\non
\\
&&=\sqrt{\frac{\hbar}{2\pi i\nu(t)}}\exp\left[i\alpha_\lambda(t)
+i\frac{\nu(t)}{2\hbar}
\theta^2_c(t)+in_0\theta_c(t)\right]\non\\
&&\times\sum^{\infty}_{n=-\infty}
\exp\left[\frac{i\hbar(n-n_0)^2}{2\nu(t)}+in\left(\theta-\theta_c(t)\right)\right].
\label{result2} \eqn
From these derivations it should be noted that {\it when using the
LR method, the boundary conditions have to be carefully managed,
otherwise one may get an incorrect result}.

\section{A Comparison to Various Approaches}

In this section we shall show that Eq.~(\ref{result1}) and
(\ref{result2}) can be cast into the forms of Eq.(\ref{pw}) and
Eq.~(\ref{pkt}). To proceed further, let us first borrow some
parameters from Sec.~II, and notice the simple result
\begin{equation}
a(t)n_0\hbar+b(t)\theta_c(t)+c(t)=a_0n_0\hbar+b_0\theta_0+c_0.
\end{equation}
Comparing this result with Eq.~(\ref{lmbd}), we find that they are
very similar. Now let us define
\begin{equation}
\lambda\equiv a(t)n_0\hbar+b(t)\theta_c(t)+c(t).\label{lmbd1}
\end{equation}

From Eq.(\ref{lmbd1}), we get \bqn
\mu(t)&=&n_0\hbar+\nu(t)\theta_c(t),\\
\non\\
\eta(t)&=&l_c(t)+\nu(t)\theta_c(t). \eqn Moreover, using the
identity \bqn \frac{d\theta^2_c(t)}{dt}=\frac{2}{I}l_c(t)\theta_c(t)
\eqn and the identity of $\dot{\nu}$ in Eq.~(\ref{munu}), we have
\begin{equation}
\eta^2=l^2_c+I\frac{d}{dt}(\nu\theta^2_c).\label{eta2}
\end{equation}

Substituting Eq.(\ref{eta2}) into Eq.(\ref{mueta}), we get
\begin{equation}
e^{i\alpha_\lambda(t)}=\frac{e^{i\alpha_\lambda(0)}\exp
\left(-\frac{i}{\hbar}\int^t_0\frac{l^2_c(\tau)}{2I}d\tau-\frac{i\nu\theta^2_c}{2\hbar}\right)
}{\sqrt{1-\frac{\nu_0t}{I}}}.
\end{equation}
Moreover, defining $e^{i\alpha_\lambda(0)}$ and $\nu_0$ as
\begin{equation}
e^{i\alpha_\lambda(0)}\equiv
\frac{N\sqrt{\pi}}{\sigma},\;\;\;\;\nu_0\equiv-\frac{iI}{T}=-\frac{i\hbar}{2\sigma^2},
\end{equation}
we see clearly that Eqs.~(\ref{result1}) and (\ref{result2}) become
exactly the same as Eqs.~(\ref{pw}) and (\ref{pkt}).  These two
approaches are thus verified to be equivalent.

It is now interesting to discuss the physical meanings of $l_c(t)$
and $\theta_c(t)$ we have obtained.  Although they are originated
from the classical treatment, however, in what sense do they play a
role of dynamic variables in the corresponding classical system
should be further clarified.  We would like to bring attention that
in the Schr\"{o}dinger picture within coordinate representation, the
$l_c(t)$ is the expectation value of
$\hat{L}-qRA(t)=-i\hbar\partial/\partial\theta-qRA(t)$. However,
$\theta_c(t)$ is not the expectation value of $\hat{\theta}=\theta$
with respect to the wave function obtained in Eq.(\ref{pw}),
instead, it is merely the peak position of the wave packet [see
Eq.(\ref{pkt})].

In other words, the conventional Ehrenfest Theorem is not directly
applicable in this topologically nontrivial system. This consequence
is due to the fact that we are not able to distinguish the phase
between the angle $\theta$ and $\theta+2n\pi$. Hence the
$\hat{\theta}$ operator is not well-defined, only the
$e^{i\hat{\theta}}$ is a well-defined operator, as has been
demonstrated in Sec.~IV. These facts cause the $\lambda$ losing its
meaning as an expectation value of the $\hat{Q}$ operator.\\

\section{SUMMARY}

In this article, we have studied the problem of a charged particle moving
in a ring subject to a time-dependent flux threading it. After
analyzing the problem in a classical manner, various approaches including
Fourier expansion method, time-evolution
method, and Lewis-Riesenfeld method are considered and compared.
In the Lewis-Riesenfeld approach, by appropriately
managing the periodic boundary condition of the system, a time-dependent wave
function can be obtained by using a non-Hermitian time-dependent linear invariant.
The eigenfunction of the invariant can be realized as a
Gaussian-type wave packet with the peak moving along the
classical angular trajectory, while the distribution of the wave
packet is determined by the ratio of the coefficient of the initial
angle to that of the initial canonical angular momentum. In this
circular system, we find that although the classical
trajectory and angular momentum can determine the motion of the wave packet;
however, the peak position is no longer
an expectation value of the angle operator, and the Ehrenfest theorem
is not directly applicable.

\sect{Acknowledgment}
The authors are grateful to D.~H. Lin and Y.~M. Kao for discussion
of the results. This work was supported partly by the National
Science Council, the National Central University, and the National
Center for Theoretical Sciences in Taiwan.

\end{document}